\documentclass{birkmult}
 \newtheorem{thm}{Theorem}[section]
 \newtheorem{cor}[thm]{Corollary}
 \newtheorem{lem}[thm]{Lemma}
 
 \theoremstyle{definition}
 \newtheorem{definition}[thm]{Definition}
 \theoremstyle{remark}
 \newtheorem{remark}[thm]{Remark}
 \newtheorem*{ex}{Example}
 \numberwithin{equation}{section}

\usepackage{graphicx}

\begin{document}
%
%
%
%
%
%
%
%
%
\title[$q$-CLT Consistent with Nonextensive Statistical Mechanics]{A generalization of the central limit theorem consistent with nonextensive statistical mechanics}
\author{Sabir Umarov}
\address{
Department of Mathematics, Tufts University\\
Medford, MA 02155\\
USA}
\email{Sabir.Umarov@tufts.edu}

\author{Constantino Tsallis}
\address{Centro Brasileiro de Pesquisas Fisicas \\
Xavier Sigaud 150, 22290-180 Rio de Janeiro-RJ \\
Brazil}

\address{
Santa Fe Institute\\
1399 Hyde Park Road, Santa Fe, NM 87501\\
USA
}
\email{tsallis@cbpf.br}

\author{Stanly Steinberg}
\address{
Department of Mathematics and Statistics\\
University of New Mexico, Albuquerque, NM 87131\\
USA}
\email{stanly@math.unm.edu}




\begin{abstract}
The standard central limit theorem plays a
fundamental role in Boltzmann-Gibbs statistical mechanics. This
important physical theory has been generalized
\cite{Tsallis1988} in 1988 by using the entropy $S_q =
\frac{1-\sum_i p_i^q}{q-1}$ (with $q \in \mathcal{R}$) instead of its
particular BG case $S_1=S_{BG}= -\sum_i p_i \ln p_i$. The theory
which emerges is usually referred to as {\it nonextensive
statistical mechanics} and recovers the standard theory for $q=1$.
During the last two decades, this $q$-generalized statistical
mechanics has been successfully applied to a considerable amount of
physically interesting complex phenomena. A conjecture\cite{Tsallis2005} and numerical
indications available in the literature have been, for a few years,
suggesting the possibility of $q$-versions of the standard
central limit theorem by allowing the random variables that are
being summed to be strongly correlated in some special manner, the
case $q=1$ corresponding to standard probabilistic independence.
This is what we prove in the present paper for $1 \leq
q<3$. The attractor, in the usual sense of a central limit theorem,
is given by a distribution of the form $p(x) =C_q [1-(1-q) \beta
x^2]^{1/(1-q)}$ with $\beta>0$, and normalizing constant $C_q$.
These distributions, sometimes referred to as $q$-Gaussians, are
known to make, under appropriate constraints, extremal the
functional $S_q$ (in its continuous version). Their $q=1$ and $q=2$ particular cases recover
respectively Gaussian and Cauchy distributions.
\end{abstract}

\maketitle

\section{INTRODUCTION}

Limit theorems and, particularly, the central limit theorems (CLT),
surely are among the most important theorems in probability theory
and statistics. They play an essential role in various applied
sciences, including statistical mechanics. Historically A. de
Moivre, P.S. de Laplace, S.D. Poisson and C.F. Gauss have first
shown that Gaussian is the attractor of independent systems with a
finite second variance. Chebyshev, Markov, Liapounov, Feller,
Lindeberg, L\'evy have contributed essentially to the development of
the central limit theorem.
\par
It is well known in the classical Boltzmann-Gibbs (BG) statistical
mechanics that the Gaussian maximizes, under appropriate
constraints, the Boltzmann-Gibbs entropy $S_{BG}=-\sum_i p_i \ln
p_i$ ($S_{BG}=-\int  dx \, p(x) \ln p(x)$ in its continuous form). The $q$-generalization of the classic entropy introduced in
\cite{Tsallis1988} as the basis for generalizing the BG theory, and
given by $S_q=\frac{1-\sum_i p_i^q}{q-1}$ ($S_{q}=\Bigl[ 1-\int  dx \, [p(x)]^q \Bigr]/(q-1)$ in its continuous form), with
$q\in \mathcal{R}$ and $S_1 =
S_{BG}$, reaches its maximum at the distributions usually referred
to as $q$-Gaussians (see \cite{PratoTsallis1999}). This fact, and a
number of conjectures \cite{Tsallis2005}, numerical indications
\cite{MoyanoTsallisGellmann2006}, and the content of some other
studies \cite{Tsallis1988,PratoTsallis1999,variousCLT,TsallisBukman}
suggest the existence of a $q$-analog of the CLT as well. In this
paper we prove a generalization of the CLT for $1 \leq q < 3.$ The
case $q<1$ requires essentially different technique, therefore we
leave it for a separate paper.
\par
In the classical CLT, the random variables are required to be independent.
Central limit theorems were established for weakly dependent random variables also.
An introduction to this area can be
found in \cite{Yoshihara1992,Peligrad1986,Rio2000,Doukhan1994,DehlingDenkerPhilipp1986,Bradley2003}
(see also references therein), where different
types of dependence are considered, as well as the history of the developments. The CLT does not hold if correlation
between far-ranging random variables is not neglectable (see \cite{DehlingMikoschSorensen2002}).  Nonextensive statistical
mechanics deals with  strongly
correlated random variables, whose correlation does {\it not} rapidly decrease with increasing  'distance' between random variables localized or moving in some geometrical lattice (or continuous space) on which a 'distance' can be defined.
This type of correlation is sometimes referred to as {\it global} correlation (see \cite{TsallisGellmannSato} for more details).
\par
In general, $q$-CLT is untractable if we rely
on the classic algebra. However, nonextensive statistical mechanics
uses constructions based on a special algebra sometimes referred to
as $q$-algebra, which depends on parameter $q$. We show that, in the
framework of $q$-algebra, the corresponding $q$-generalization of
the central limit theorem becomes possible and relatively simple.
\par

The theorems obtained in this paper are represented as a series of
theorems depending on the type of correlations. We will consider
three types of correlation. An important distinction from the
classic CLT is the fact that a complete formulation of $q$-CLT is not possible
within only one given $q$. The parameter $q$ is connected with two
other numbers, $q_{\ast}= z^{-1}(q) \, \mbox{and} \,
q^{\ast}=z(q),$ where $z(s)=(1+s)/(3-s).$ We will see that
$q_{\ast}$ identifies an attractor, while $q^{\ast}$ yields
the scaling rate. In general, the present $q$-generalization of
the CLT is connected with a triplet $(q_{k-1}, \,
q_k, \, q_{k+1})$ determined by a given $q \in [1,2)$. For systems
having correlation identified by $q_k$, the index $q_{k-1}$
determines the attracting $q$-Gaussian, while the index $q_{k+1}$
indicates the scaling rate. Note that, if $q=1$, then the entire
family of theorems reduces to one element, thus recovering the
classic CLT.
\par

The paper is organized as follows. In Section 2 we recall the basics
of $q$-algebra, definitions of {\it $q$-exponential} and {\it
$q$-logarithm}. Then we introduce a transform $F_q$ and study its
basic properties. For $q \neq 1$ $F_q$ is not a linear operator.
Note that $F_q$ is linear only if $q=1$ and in this case coincides
with the classic Fourier transform.  Lemma \ref{centrallemma}
implies that $F_q$ is invertible in the class of $q$-Gaussians. An
important property of $F_q$ is that it maps a $q$-Gaussian to a
$q^{\ast}$-Gaussian (with a constant factor). In Section 3 we prove
the main result of this paper, i.e., the $q$-version of the CLT. We introduce  the notion of $q$-independent random variables (three types), which classify  correlated
random variables. Only in the case $q=1$ the correlation disappears,
thus recovering the classic notion of independence of random
variables.

\section{$q$-ALGEBRA and $q$-FOURIER TRANSFORM}

\subsection{$q$-sum and $q$-product}
The basic operations of the $q$-algebra appear naturally in
nonextensive statistical mechanics. It is well known that, if $A$ and
$B$ are two independent subsystems, then the total BG entropy
satisfies the additivity property
\[
\label{additivity} S_{BG}(A+B)=S_{BG}(A)+S_{BG}(B).
\]
Additivity is not preserved for $q\ne 1$. Indeed, we easily verify
\cite{Tsallis1988,PratoTsallis1999,Tsallis2005}
\[
S_q(A+B)=S_q(A)+S_q(B)+(1-q)\,S_q(A)\,S_q(B).
\]
Introduce the $q$-sum of two numbers $x$ and $y$ by the formula 
$$
x \oplus_q y = x+y+(1-q)xy.
$$ 
Then, obviously, $S_q(A+B)=S_q(A)
\oplus_q S_q(B).$ It is readily seen that the $q$-sum is
commutative, associative, recovers the usual summing operation if
$q=1$ (i.e. $x \oplus_1 y = x+y$), and preserves $0$ as the neutral
element (i.e. $x \oplus_q 0 = x$). By inversion,
we can define the {\it $q$-subtraction} as $x \ominus_q y =
\frac{x-y}{1+(1-q)y}.$ Further, the {\it $q$-product} for $x,y$ is
defined by the binary relation $x \otimes_q y =
[x^{1-q}+y^{1-q}-1]_+^{{1}\over{1-q}}$. The symbol $[x]_+$ means that
$[x]_+ = x$, if $x \geq 0$, and $[x]_+ = 0$, if $x \le 0$.
Also this operation is
commutative, associative, recovers the usual product when $q=1$ (i.e. $x \otimes_1 y=xy$),
preserves $1$ as the unity (i.e. $x \otimes_q 1=x$).
Again, by inversion, a {\it $q$-division} can be defined: $x
\oslash_q y = [x^{1-q}-y^{1-q}+1]_+^{1 \over {1-q}}$.
Note that,
for $q \neq 1$, division by zero is
allowed.  As we will see below, the $q$-sum and the $q$-product
are connected each other through the $q$-exponential, generalizing
the fundamental property of the exponential function, $e^{x+y}=e^x
e^y$ (or $\ln (xy)=\ln x +\ln y$ in terms of logarithm).
\par
\subsection{$q$-exponential and $q$-logarithm}
The $q$-analysis relies essentially on the analogs of exponential
and logarithmic functions, which are called $q$-exponential and
$q$-logarithm \cite{TsallisQuimicaNova}. In the mathematical
literature there are other generalizations of the classic
exponential distinct from the $q$-exponential used in the present
paper. These generalizations were introduced by Euler \cite{Euler},
Jackson \cite{Jackson}, and others. See \cite{Ernst} for details.

The {\it $q$-exponential} and {\it $q$-logarithm}, which are denoted by $e_q^x$
and $ln_q \,x,$ are respectively defined as
$e_q^x=[1+(1-q)x]_+^{{1}\over {1-q}}$ and $\ln_q x=
\frac{x^{1-q}-1}{1-q}, \, (x>0).$

For the $q$-exponential, the relations $e_q^{x \oplus_q y} = e_q^x e_q^y$
and $e_q^{x+y}=e_q^x \otimes_q e_q^y$ hold true. These relations can
be rewritten equivalently as follows: $\ln_q (x \otimes_q y)=\ln_q x +
\ln_q y$, and $\ln_q (x y)=\ln_q x \oplus_q \ln_q y$\footnote{These
properties reflect the possible extensivity of the nonadditive entropy $S_q$ in the presence
of special correlations
\cite{TsallisGellmannSato,MarshallEarl,tsallisEPN,MarshFuentesMoyanoTsallis,CarusoTsallis}.}.
The $q$-exponential and $q$-logarithm have asymptotics $e_q^x = 1 +
x + {q \over 2}x^2 + o(x^2), \, x \rightarrow 0$ and $\ln_q (1+x) =
x - {q \over 2} x^2 + o(x^2), \, x \rightarrow 0.$ For $q \neq 1$, we
can define $e_q^{iy}$, where $i$ is the imaginary unit and $y$ is
real, as the principal value of $[1+i(1-q)y]^{{1}\over {1-q}}$,
namely
\[
e_q^{iy} = [1+(1-q)^2 y^2]^{\frac{1}{2(1-q)}} e^{\frac{i
\arctan[(q-1)y]}{1-q}}, \, \, q \neq 1.
\]
Analogously, for $z=x+iy$ we can define $e_q^z= e_q^{x+iy},$ which is
equal to $e_q^x \otimes_q e_q^{iy}$ in accordance with the property
mentioned above. Note that, if $q < 1,$ then for real $y$,
$|e_q^{iy}| \geq 1$ and $|e_q^{iy}| \sim (1+y^2)^{\frac{1}{2(1-q)}},
\, x \rightarrow \infty.$ Similarly, if $q > 1$, then $0 <
|e_q^{iy}| \leq 1$ and $|e_q^{iy}| \rightarrow 0$ if $|y|
\rightarrow \infty.$
\subsection{$q$-Gaussian}
Let $\beta$ be a positive number. We call the function
\begin{equation}
\label{gaussian}
G_q(\beta; x)= \frac{\sqrt{\beta}}{C_q } e_q^{- \beta x^2} \,.
\end{equation}
a {\it $q$-Gaussian}; $C_q$ is the normalizing constant, namely
$C_q = \int_{-\infty}^{\infty}e_q^{-x^2}dx.$
It is easy to verify that
\begin{equation}
C_q =
\left\{\begin{array}{ll}
   {
\frac{2}{\sqrt{1-q}} \int_0^{\pi / 2} (\cos \, t)^{\frac{3-q}{1-q}} dt }
= \frac{2\sqrt{\pi}\, \Gamma\bigl({1 \over {1-q}}\bigr)}
{(3-q) \sqrt{1-q} \, \Gamma\bigl({{3-q} \over {2(1-q)}}\bigr)}, & -\infty<q<1, \\

{\sqrt{\pi}}, & q=1, \\

\frac{2}{\sqrt{q-1}} \int_0^{\infty} (1+y^2)^{{-1} \over {q-1}} dy =
\frac{\sqrt{\pi} \, \Gamma\bigl(\frac{3-q}{2(q-1)}\bigr)}{\sqrt{q-1}
\, \Gamma \bigl({1 \over {q-1}}\bigr)}
, &  1<q<3 \,. \\
\end{array} \right.
\end{equation}
For $q<1$, the support of $G_q(\beta;x)$ is compact since this density vanishes for $|x| > 1/\sqrt{(1-q)\beta}$.
Notice also that, for $q < 5/3$ ($5/3 \le q <3$), the variance is finite (diverges).
Finally, we can easily check that there are relationships between
different values of $q$. For example,
$e_q^{-x^2} =  \Bigl[ e_{2-{1 \over q}}^{-q x^2} \Bigr]^{1 \over q}.$
\par
The following lemma establishes a general relationship (which contains the previous one as a particular case) between different
$q$-Gaussians.
\begin{lem}
\label{l1}
For any real $q_1$, $\beta_1 >0$ and $\delta >0$ there exist
uniquely determined
$q_2=q_2(q_1, \delta)$ and $\beta_2 = \beta_2(\delta, \beta_1),$ such
that
\[
(e_{q_1}^{-\beta_1 x^2})^{\delta} = e_{q_2}^{-\beta_2 x^2}.
\]
Moreover, $q_2=\delta^{-1}(\delta - 1 +q_1),$ $\beta_2 = \delta
\beta_1.$
\end{lem}
\begin{proof}
Let $q_1 \in R^1, \beta_1 >0$ and $\delta >0$ be any fixed real
numbers. For the equation
\[
(1 - (1-q_1)\beta_1 x^2)^{\delta \over {1-q_1}} = (1- (1-q_2) \beta_2
x^2)^{1 \over {1-q_2}}
\]
to be an identity, it is needed $(1-q_1)\beta_1 = (1-q_2)\beta_2,$
$1-q_1 = \delta (1-q_2).$
These equations have a unique solution $q_2=\delta^{-1}(\delta - 1
+q_1),$
$\beta_2 = \delta \beta_1.$
\end{proof}
\par
The set of all $q$-Gaussians with a positive constant factor will be
denoted by $\mathcal{G}_q $, i.e.,
$$
\mathcal{G}_q = \{b \,G_q(\beta, x): b>0, \beta > 0\}.
$$
\subsection{$q$-Fourier transform}
From now on we assume that $1 \leq q <3.$ For these values of $q$ we
introduce the {\it $q$-Fourier transform} $F_q$,
an operator which coincides with the Fourier transform if $q=1$.  Note that the
$q$-Fourier transform is defined on the basis of the $q$-product and the
$q$-exponential, and, in contrast to the usual Fourier transform, is
a {\it nonlinear} transform for $q\in(1,3)$. Let $f$ be a non-negative
measurable function with $supp \, f \subseteq R.$ The $q$-Fourier
transform for $f$ is defined by the formula
\begin{equation}
\label{Fourier} F_q[f](\xi) = \int_{supp \, f} e_q^{ix\xi} \otimes_q
f(x) dx \,,
\end{equation}
where the integral is understood in the Lebesgue sense.
For discrete functions $f_k, k \in \mathcal{Z}= \{0, \pm 1,...\},$
$F_q$ is defined as
\begin{equation}
\label{FourierDiscrete}
F_q[f](\xi) = \sum_{k \in \mathcal{Z}_f} e_q^{ik\xi} \otimes_q f_k \,,
\end{equation}
where $\mathcal{Z}_f = \{k \in \mathcal{Z}: f_k \neq 0\}.$

In what follows we use the same notation in both cases. We also call
(\ref{Fourier}) or (\ref{FourierDiscrete}) the {\it
$q$-characteristic function} of a given random variable $X$ with an
associated density $f(x),$ using the notations $F_q(X)$ or $F_q(f)$
equivalently. The following lemma establishes the expression of the
$q$-Fourier transform in terms of the standard product, instead of the $q$-product.
\begin{lem}
\label{informal}
The $q$-Fourier transform can be written in the form
\begin{equation}
\label{identity2} F_q[f](\xi) = \int_{-\infty}^{\infty}f(x)\, e_q^{ix
\xi [f(x)]^{q-1}}dx.
\end{equation}
\end{lem}
\vspace{.2cm}

\begin{proof} 
For $x \in supp \, f$ we have
\[
e_q^{ix \xi} \otimes_q f(x)=[1 + (1-q)ix \xi + [f(x)]^{1-q}-1]^{1
\over {1-q}} =
\]
\begin{equation}
\label{l2} f(x)[1 + (1-q) ix \xi [f(x)]^{q-1}]^{1 \over {1-q}}.
\end{equation}
Integrating both sides of Eq. (\ref{l2}) we obtain (\ref{identity2}).
\end{proof}

Analogously, for a discrete $f_k, \, k \in \mathcal{Z},$ (\ref{FourierDiscrete}) can be represented as
\[
F_q[f](\xi) = \sum_{k \in \mathcal{Z}} f_k e_q^{ik\xi f_k^{q-1}} \,.
\]
\begin{cor}
\label{1corinformal} The $q$-Fourier transform exists for any
nonnegative $f \in L_1 (R).$
Moreover, $|F_q[f](\xi)| \leq \|f\|_{L_1}.$ \footnote{Here, and
elsewhere, $\|f\|_{L_1} = \int_{\mathcal{R}} f(x) dx$, $L_1$ being
the space of absolutely integrable functions.}
\end{cor}
\begin{proof}
This is a simple implication of Lemma \ref{informal} and of the asymptotics of
$e_q^{ix}$ for large $|x|$ mentioned above.
\end{proof}
\begin{cor}
\label{uniqueness}
Assume $f (x) \geq 0, \, x \in R$ and
$F_q[f] (\xi) = 0$ for all $\xi \in R.$ Then $f(x) = 0$ for almost all $x \in R.$
\end{cor}
\begin{lem}
\label{centrallemma} Let $1 \leq q < 3.$ For the $q$-Fourier
transform of a $q$-Gaussian, the following formula holds:
\begin{equation}
\label{gausstransform0}
F_q[G_q(\beta; x)](\xi) = \Bigl(e_q^{\, -\frac{\xi^2}{4 \beta^{2-q}
C_q^{2(q-1)}}}\Bigr)^{{3-q \over 2}}.
\end{equation}
\end{lem}
\begin{proof}
Denote $a=\frac{\sqrt{\beta}}{C_q}$ and write
\[
F_q[a \,e_q^{-\beta x^2}](\xi) = \int_{-\infty}^{\infty} (a \,e_q^{-\beta
x^2})\otimes_q (e_q^{ix \xi}) dx
\]
using the property $e_q^{x+y}=e_q^x \otimes_q e_q^y $
of the $q$-exponential, in the form
\[
F_q[a e_q^{-\beta x^2}](\xi) = a \int_{-\infty}^{\infty} e_q^{-\beta
x^2 + i a^{q-1} x \xi } dx =
        a \int_{-\infty}^{\infty} e_q^{-(\sqrt{\beta} x - \frac{i
a^{q-1} \xi}{2 \sqrt{\beta}})^2 - \frac{a^{2(q-1)} \xi^2}{4 \beta}} dx =
\]
\[
a \int_{-\infty}^{\infty} e_q^{-(\sqrt{\beta} x - \frac{i a^{q-1}
\xi}{2 \sqrt{\beta}})^2} \otimes_q e_q^{ - \frac{a^{2(q-1)} \xi^2}{4 \beta}}
dx.
\]
The substitution $y=\sqrt{\beta} x - \frac{i a^{q-1}\xi}{2
\sqrt{\beta}}$ yields the equation
\[
F_q[a e_q^{-\beta x^2}](\xi)=
\frac{a}{\sqrt{\beta}}  \int_{-\infty+i \eta}^{\infty +i \eta}
e_q^{-y^2} \otimes_q e_q^{ - \frac{a^{2(q-1)} \xi^2}{4 \beta}} dy \,,
\]
where $\eta=\frac{\xi a^{q-1}}{2 \sqrt{\beta}}$. Moreover, the
Cauchy theorem on integrals over closed curves is applicable
because of at least a power-law decay of $q$-exponential for any $1
\leq q<3$. By using it, we can transfer the integration from $R + i\eta$ to R.
Hence, applying again Lemma \ref{informal}, we have
\[
F_q[G_q(\beta; x)](\xi) = \frac{a e_q^{-\frac{a^{2(q-1)}}{4 \beta}
\xi^2}}{\sqrt{\beta}}
\int_{-\infty}^{\infty} e_q^{   -y^2 \Bigl(   e_q^{-\frac{a^{2(q-1)}}{4
\beta} \xi^2}   \Bigr)^{q-1}      }dy=
\]
\[
\frac{a C_q}{\sqrt{\beta}} \Bigl(    e_q^{   -\frac{a^{2(q-1)} \xi^2 }{4
\beta} }   \Bigr)^{1-{{q-1} \over 2}}.
\]
Simplifying the last expression, we arrive to Eq. (\ref{gausstransform0}).
\end{proof}

Introduce the function $z(s)=\frac{1+s}{3-s}$ for $s \in (-\infty, 3),$
and denote its inverse
$z^{-1}(t), \, t \in (-1, \infty)$.
It can be easily verified that $z\bigl(\frac{1}{z(s)}\bigr)= {1 \over s}$ and
$z({1 \over s})={1 \over z^{-1}(s)}\,.$
Let $q_{1}=z(q)$ and $q_{-1}=z^{-1}(q)\,.$
It follows from the mentioned properties of $z(q)$ that
\begin{equation}
\label{qdual1}
z\Bigl({1 \over q_1}\Bigr)={1 \over q} \, \, \, \, \, \mbox{and} \, \, \, \, \, z\Bigl({1 \over q}\Bigr)={1 \over q_{-1}}.
\end{equation}
The function $z(s)$ also possess the following two important
properties
\begin{equation}
\label{qdual2}
z(s) \, z(2-s)=1 \, \, \, \, \, \mbox{and} \, \, \, \, \, z(2-s) + z^{-1} (s) = 2.
\end{equation}
It follows from these
properties that $q_{-1}+ {1 \over q_1} =2\,.$
\begin{cor}
\label{centrallemmacor}
For $q$-Gaussians the following $q$-Fourier transforms hold
\begin{equation}
\label{gausstransform21} F_q[G_q(\beta; x)](\xi) = e_{q_{1}}^{-
\beta_{\ast} (q) \xi^2}, \, \, q_{1} = z(q), \, 1\leq q<3;
\end{equation}
\begin{equation}
\label{gausstransform22} F_{q_{-1}}[G_{q_{-1}}(\beta; x)](\xi) =
e_{q}^{- \beta_{\ast} (q_{-1}) \xi^2}, \, \, q_{-1} = z^{-1}(q), \,
1 \leq q<3,
\end{equation}
where  $\beta_{\ast} (s) = \frac{3-s}{8 \beta^{2-s} C_s^{2(s-1)}} \,$ (or, more symmetrically, $\beta_{\ast}^{\frac{1}{\sqrt{2-s}}}\beta^{\sqrt{2-s}} =K(s)$ with $K(s)= \Bigl[ \frac{3-s}{8  C_s^{2(s-1)}}\Bigr]^{\frac{1}{\sqrt{2-s}}} $; $0 \le K(s)<1$ for $s\le 2$, with $\lim_{s\to -\infty}K(s)=K(2)=0$).
\end{cor}
\begin{remark}
Note that $\beta_{\ast} (s) >0$ if $s<3 \,.$
\end{remark}
\begin{cor}
\label{cor2.9}
The following mappings
\[
F_q:\mathcal{G}_q \rightarrow \mathcal{G}_{q_{1}} \, , q_{1} = z(q),
\, 1 \leq q<3,
\]
\[
F_{q_{-1}}: \mathcal{G}_{q_{-1}} \rightarrow \mathcal{G}_q, \, \,
q_{-1} = z^{-1}(q), \, 1 \leq q <3,
\]
hold and they are injective.
\end{cor}
\begin{cor}
There exist the following inverse $q$-Fourier transforms
\[
F^{-1}_q: \mathcal{G}_{q_{1}} \rightarrow \mathcal{G}_q,   \, \,
q_{1} = z(q), \, 1 \leq q<3,
\]
\[
F^{-1}_{q_{-1}}:  \mathcal{G}_q \rightarrow \mathcal{G}_{q_{-1}} \,
, q_{-1} = z^{-1}(q), \, 1 \leq q <3.
\]
\end{cor}
\begin{lem}
The following mappings
\[
F_{{1 \over q_{1}}}: \mathcal{G}_{{1 \over q_{1}}} \rightarrow
\mathcal{G}_{1 \over q}, \, q_{1} = z(q), \, 1 \leq q<3,
\]
\[
F_{{1 \over q}}: \mathcal{G}_{{1 \over q}} \rightarrow
\mathcal{G}_{1 \over q_{-1}}, \, \, q_{-1} = z^{-1}(q), \, 1 \leq q
< 3.
\]
hold.
\end{lem}
\begin{proof}
The assertion of this lemma follows from Corollary
\ref{cor2.9} taking into account the properties (\ref{qdual1}).
\end{proof}

Let us introduce the sequence $q_n = z_n(q) = z(z_{n-1}(q)), n=1,2,...,$  with a given $q = z_0(q), \, q<3.$
We can extend the sequence $q_n$ for negative integers $n=-1,-2,...$ as well putting
$q_{-n} = z_{-n}(q)=z^{-1}(z_{1-n}(q)), n = 1, 2,...\,.$
It is not hard to verify that
\begin{equation}
\label{qn} q_n = \frac{2q + n(1-q)}{2 + n(1-q)} = 1 +
\frac{2(q-1)}{2-n(q-1)}, \, \, n=0, \pm1, \pm 2,... \,,
\end{equation}
which, for $q \ne 1$,  can be rewritten as $\frac{2}{1-q_n}= \frac{2}{1-q}+n$. Note that $q_n \equiv 1$ for all $n=0,
\pm 1, \pm 2,...,$ if $q=1$ and $lim_{ n \rightarrow \pm
\infty}z_n(q)=1$ for all $q \neq 1.$ It follows from (\ref{qn}) that
$q_n > 1$ for all $n < 2/(q-1), \, q \in(1,3).$ Moreover, obviously,
$2/(q-1) > 1$, if $q \in(1,3)$, which implies $q_1 > 1$.
Generalizing what has been said, we can conclude that the condition
$q_n > 1$ guarantees $q_k> 1, \, k=n-1, n, n+1,$ for three
consequent members of the sequence (\ref{qn}).

Let us note also that the definition of the sequence $q_n$ can be
given through the following series of mappings.
\begin{definition}
\begin{equation}
\label{zright}
... \stackrel{z}{\rightarrow}q_{-2} \stackrel{z}{\rightarrow}q_{-1} \stackrel{z}{\rightarrow}q_{0}=q
\stackrel{z}{\rightarrow}q_{1} \stackrel{z}{\rightarrow}q_{2} \stackrel{z}{\rightarrow}...
\end{equation}
\begin{equation}
\label{zleft}
... \stackrel{z^{-1}}{\leftarrow}q_{-2}\stackrel{z^{-1}}{\leftarrow}q_{-1}\stackrel{z^{-1}}{\leftarrow}q_{0}=q \stackrel{z^{-1}}{\leftarrow}q_{1}
\stackrel{z^{-1}}{\leftarrow}q_{2} \stackrel{z}{\leftarrow} ...
\end{equation}
\end{definition}
Furthermore, we set, for $k=1,2,...$ and $n = 0, \pm 1,... \,, $
$$
F^k_{q_n}  = F_{q_{n+k-1}} \circ ... \circ F_{q_n} \,,
$$
and
\[
F^{-k}_{q_n} = F^{-1}_{q_{n-k}} \circ ... \circ F^{-1}_{q_{n-1}}\,.
\]
Additionally, for $k=0$ we let $F^0_q[f]=f.$
Summarizing the above mentioned
relationships related to $z_n(q)$, we obtain the following assertions.
\begin{lem}
\label{qnrelation}
There holds the following duality relations
\begin{equation}
\label{dualityrelation}
q_{n-1} + \frac{1}{q_{n+1}} = 2, \, n \in \mathcal{Z}.
\end{equation}
\end{lem}
\begin{proof}
Making use the properties (\ref{qdual2}), we obtain
\[
q_{n-1}=z^{-1}(q_n) = 2 - z(2-q_n)=2- \frac{1}{z(q_n)} = 2 - \frac{1}{q_{n+1}}.
\]
\end{proof}

\begin{lem}
The following mappings hold:
\[
F^k_{q_n}:\mathcal{G}_{q_n} \rightarrow \mathcal{G}_{q_{k+n}}, \, \,
\, k , \, n \in \mathcal{Z};
\]
\[
\lim_{ k \rightarrow - \infty} F^{k}_{q_n} \mathcal{G}_{q_n}  =
\mathcal{G}
\]
where $\mathcal{G}$ is the set of classical Gaussians.
\end{lem}
\begin{lem}
\label{Fqseries}
The following mappings hold:
\begin{equation}
\label{Fright}
... \stackrel{F_{q_{-3}}}{\rightarrow}\mathcal{G}_{q_{-2}} \stackrel{F_{q_{-2}}}{\rightarrow} \mathcal{G}_{q_{-1}}
\stackrel{F_{q_{-2}}}{\rightarrow} \mathcal{G}_{q} \stackrel{F_q}{\rightarrow} \mathcal{G}_{q_{1}}
\stackrel{F_{q_1}}{\rightarrow} \mathcal{G}_{q_{2}}  \stackrel{F_{q_2}}{\rightarrow} ...
\end{equation}
\begin{equation}
\label{Fleft}
... \stackrel{F^{-1}_{q_{-3}}}{\leftarrow} \mathcal{G}_{q_{-2}} \stackrel{F^{-1}_{q_{-2}}}{\leftarrow} \mathcal{G}_{q_{-1}}
\stackrel{F^{-1}_{q_{-1}}}{\leftarrow} \mathcal{G}_{q}  \stackrel{F^{-1}_q}{\leftarrow} \mathcal{G}_{q_{1}}
\stackrel{F^{-1}_{q_1}}{\leftarrow} \mathcal{G}_{q_{2}}  \stackrel{F^{-1}_{q_2}}{\leftarrow} ...
\end{equation}
\end{lem}
Note that right sides of sequences in (\ref{Fright}) and
(\ref{Fleft}) cut off for $n \geq 2/(q-1).$
%
\section{$q$-GENERALIZATION OF THE CENTRAL LIMIT THEOREM}
\subsection{$q$-independent random variables}
In this section we establish a $q$-generalization of the classical
CLT (see, e.g. \cite{Billingsley,Durrett}) for independent
identically distributed random variables with a finite variance.
First we introduce some notions necessary to formulate the
corresponding results. Let $(\Omega, \mathcal{F}, P)$ be a
probability space and $X$ be a random variable defined on it with a
density $f \in L_q(R), \, \nu_q(f)= \|f\|_{L_q}^q =
\int_{-\infty}^{\infty} [f(x)]^q dx < \infty.$ Introduce the density
$$f_q(x)=\frac{[f(x)]^q}{\nu_q(f)},$$ which is commonly referred to as
the {\it escort density} \cite{Beckbook}. Further, introduce for $X$
the notions of {\it $q$-mean}, $\mu_q = \mu_q(X) =
\int_{-\infty}^{\infty} x f_q(x)\,dx,$ and {\it $q$-variance}
$\sigma^2_q = \sigma^2_q(X - \mu_q) = \int_{-\infty}^{\infty}
(x-\mu_q)^2 f_q(x)\,dx,$
subject to the integrals used in these definitions to converge.

The formulas below can be verified directly.
\begin{lem}
\label{qmeanproperties}
The following formulas hold true
\begin{enumerate}
\item
$\mu_q (aX)=a \mu_q(X);$
\item
$\mu_q(X- \mu_q (X))= 0;$
\item
$\sigma_{q}^2(aX) = a^{2}\sigma_q^2 (X).$
\end{enumerate}
\end{lem}
Further, we introduce the notions of {\it $q$-independence} and {\it $q$-convergence}.
\begin{definition}
Two random variables $X_1$ and $Y_1$ are said
\begin{enumerate}
\item
$q$-independent of first type if
\begin{equation}
\label{q-independence1} F_{q}[X+Y](\xi)=F_q[X](\xi) \otimes_{q}
F_q[Y](\xi),
\end{equation}
\item
$q$-independent of second type if
\begin{equation}
\label{q-independence2} F_{q_{-1}}[X+Y](\xi)=F_{q_{-1}}[X](\xi)
\otimes_{q} F_{q_{-1}}[Y](\xi), \, \, q = z(q_{-1}),
\end{equation}
\item
$q$-independent of third type if
\begin{equation}
\label{q-independence3} F_{q_{-1}}[X+Y](\xi)=F_q[X](\xi) \otimes_{q}
F_q[Y](\xi), \, \, q=z(q_{-1}),
\end{equation}
where $X=X_1-\mu_q(X_1)$, \, $Y=Y_1-\mu_q(Y_1).$
\end{enumerate}
\end{definition}
All three types of $q$-independence generalize the classic notion of
independence. Namely, for $q=1$ the conditions
(\ref{q-independence1})-(\ref{q-independence3}) turn into the well
known relation
\[
F[f_X \ast f_Y] = F[f_X] \cdot F[f_Y]
\]
between the convolution (noted $\ast$) of two densities and the
multiplication of their (classical) Fourier images, and holds for
independent $X$ and $Y$. If $q \neq 1$, then $q$-independence of a
certain type describes a specific correlation. The relations
(\ref{q-independence1})-(\ref{q-independence3}) can be rewritten in
terms of densities. Let $f_X$ and $f_Y$ be densities of $X$ and $Y$
respectively, and let $f_{X+Y}$ be the density of $X+Y$. Then, for
instance, the $q$-independence of second type takes the form
\begin{equation}
\label{q-independence20} \int_{-\infty}^{\infty}e_{q_{-1}}^{i x \xi}
\otimes_{q_{-1}} f_{X+Y}(x) dx = F_{q_{-1}}[f_X](\xi) \otimes_{q}
F_{q_{-1}}[f_Y](\xi).
\end{equation}
Consider an example of $q$-independence of second type. Let random
variables $X$ and $Y$ have $q$-Gaussian densities
$G_{q_{-1}}(\beta_1, x)$ and $G_{q_{-1}}(\beta_2, x)$ respectively.
Denote $\gamma_j = \frac{3-q_{-1}}{ 8 \beta^{2-q_{-1}}_j
C_{q_{-1}}^{2(q_{-1}-1)} }, \, j=1,2.$ If $X+Y$ is distributed
according to the density $G_{q_{-1}}(\delta, x)$, where $\delta =
\left( \frac{3-q_{-1}}{ 8 (\gamma_1 + \gamma_2)
C_{q_{-1}}^{2(q_{-1}-1)}} \right)^{\frac{1}{2-q_{-1}}} $, then
(\ref{q-independence20}) is satisfied. Hence, $X$ and $Y$ are
$q$-independent of second type.

The reader can easily modify the definition of $q$-independence to
the more general case of $q_{k}$-independence. For example,
relation (\ref{q-independence2}) in the case of $q_k$-independence
takes the form
\begin{equation}
\label{q-independence2k} F_{q_{k-1}}[X+Y](\xi)=F_{q_{k-1}}[X](\xi)
\otimes_{q_k} F_{q_{k-1}}[Y](\xi), \, \, q_k = z(q_{k-1}).
\end{equation}

\begin{definition}
Let $X_1, X_2,...,X_N,...$ be a sequence of identically distributed
random variables. Denote $Y_N=X_1+...+X_N$. By definition, $X_N,
N=1,2,...$ is said to be $q_{k}$-independent (or $q_k$-i.i.d.) of
first type if, for all $N=2,3,...$, the relations
\begin{equation}
\label{qiid1} F_{q_{k}}[Y_N - N \mu_k](\xi)=F_{q_k}[X_1-\mu_k](\xi)
\otimes_{q_{k}} ... \otimes_{q_{k}} F_{q_k}[X_N-\mu_k](\xi), \,
\mu_k =\mu_{q_k}(X_1),
\end{equation}
hold.
\end{definition}
Analogously, $q_k$-independence of second and third types can be
defined for sequences of identically distributed random variables.

\begin{remark}
For $k=-1$ it follows from this definition the $q$-independence of a sequence of random variables, namely
\begin{equation}
\label{q0iid}
F_{q_{-1}}[Y_N-N \mu_{-1}](\xi)=F_{q_{-1}}[X_1-\mu_{-1}](\xi) \otimes_{q} ... \otimes_{q} F_{q_{-1}}[X_N-\mu_{-1}](\xi), \, \, N=2,3,...
\end{equation}
\end{remark}
\begin{ex}
Assume that $X_N, N=1,2,...,$ is the sequence of identically distributed random variables
with the associated density $G_{q_{-1}}(\beta, x)$. Further, assume the sums $X_1+...+X_N, \, N=2,3,...,$ are distributed according
to the density $G_{q_{-1}}(\alpha, x)$, where $\alpha = N^{-\frac{1}{2-q_{-1}}} \beta$. Then the sequence $X_N$ satifies
the relation (\ref{q0iid}) for all $N=2,3,...$, thus being a $q$-independent identically distributed sequence of random variables.
\end{ex}

\begin{definition}
\label{qconvergence} A sequence of random variables $X_N, \,
N=1,2,...,$ is said to be $q$-convergent to a random variable
$X_{\infty}$ if $\lim_{N \rightarrow \infty} F_q [X_N](\xi) = F_q
[X_{\infty}](\xi)$ locally uniformly in $\xi$.
\end{definition}

Evidently, this definition is equivalent to the weak convergence of
random variables if $q=1.$ For $q \neq 1$ denote by $W_q$ the set of
continuous functions $\phi$ satisfying the condition $|\phi(x)| \leq
C(1+|x|)^{-\frac{q}{q-1}}, \, x \in R$.

\begin{definition}
\label{qweakconvergence} A sequence of random variables $X_N$ with
the density $f_N$ is called weakly $q$-convergent to a random
variable $X_{\infty}$ with the density $f$ if $\int_{R} f_N(x) d m_q
\rightarrow \int_{R^d} f(x)d m_q $ for arbitrary measure $m_q$
defined as $dm_q(x)= \phi_q (x) dx,$ where $\phi_q \in W_q$.
\end{definition}

The $q$-weak convergence is equivalent to the $q$-convergence
\cite{UmarovTsallis2007}.

We will study limits of sums
\[
Z_N = \frac{1}{D_{N}(q)} \, (X_1 + ...+ X_N -N \mu_q), N=1,2,...
\]
where $D_{N}(q)=(\sqrt{N \nu_{2q-1}} \sigma_{2q-1})^{\frac{1}{2-q}},
\, N=1,2,...,$ in the sense of Definition \ref{qconvergence}, when
$N \rightarrow \infty$. Namely, the question we are interested in:
{\it Is there a $q$-normal distribution that attracts the sequence
$Z_N$? } For $q=1$ the answer is well known and it is the content of
the classical central limit theorem.

\subsection{Main results}

The formulation of a generalization of the central limit theorem
consistent with nonextensive statistical mechanics depends on the
type of $q$-independence. We prove the $q$-generalization of the
central limit theorem under the condition of first type of
$q$-independence.

\vspace{.3cm}

{\bf Theorem 1.} {\it Assume a sequence $q_k, k \in \mathcal{Z},$ is
given as (\ref{qn}) with $q_{k} \in[1,2).$ Let $X_1, ..., X_N,...$
be a sequence of $q_k$-independent (for a fixed $k$) of first type
and identically distributed random variables with a finite
$q_{k}$-mean $\mu_{q_{k}}$ and a finite second $(2q_{k}-1)$-moment
$\sigma_{2q_{k}-1}^2$ .

Then $Z_N = \frac{X_1 + ... + X_N -N \mu_{q_{k}}}{D_{N}(q_{k})}$
is $q_{k}$-convergent to a $q_{k-1}$-normal distribution as $N
\rightarrow \infty$. Moreover, the corresponding attractor is
$G_{q_{k-1}}(\beta_k; x)$ with
\begin{equation}
\label{betak}
\beta_k = \Bigl(\frac{3-q_{k-1}}{4 q_{k}
C_{q_{k-1}}^{2 q_{k-1} -2}}\Bigr)^{1 \over {2-q_{k-1}}}.
\end{equation}}
The proof of this theorem follows from Theorem 2 proved below and
Lemma \ref{Fqseries}. Theorem 2 represents one element ($k=0$) in
the series of assertions contained in Theorem 1.
\vspace{.3cm}

{\bf Theorem 2.} {\it Assume $1 \leq q < 2$. Let $X_1, ..., X_N,...$
be a sequence of $q$-independent of first type and identically
distributed random variables with a finite $q$-mean $\mu_q$ and a
finite second $(2q-1)$-moment $\sigma_{2q-1}^2.$
\par
Then $Z_N = \frac{X_1 + ... + X_N -N \mu_{q}}{D_{N}(q)}$ is
$q$-convergent to a $q_{-1}$-normal distribution as $N \rightarrow
\infty$. The corresponding $q_{-1}$-Gaussian is 
$G_{q_{-1}}(\beta; x),$ with  
$$\beta = \Bigl(\frac{3-q_{-1}}{4 q C_{q_{-1}}^{2 q_{-1}
-2}} \Bigr)^{1 \over {2-q_{-1}}}.$$ }
\begin{proof}
Let $f$ be the density associated with $X_1-\mu_{q}$.
First we evaluate $F_{q}(X_1-\mu_{q})=F_{q}(f(x)).$ Using Lemma
\ref{informal} we have
\begin{equation}
\label{step_10} F_{q} [f](\xi) = \int_{-\infty}^{\infty}e_{q}^{ix
\xi} \otimes_{q} f(x) \, dx =
                              \int_{-\infty}^{\infty} f(x) \,
e_{q}^{ix \xi[f(x)]^{q-1}}dx .
\end{equation}
Making use of the asymptotic expansion
$e_q^x = 1 + x + {q \over 2}x^2 + o(x^2), \, x \rightarrow 0,$
we can rewrite the right hand side of (\ref{step_10})
in the form
\vspace{.2cm}

\hspace{1cm}$F_{q} [f](\xi) =  $
$$
   \int_{-\infty}^{\infty} f(x) \left(1 + ix  \xi [f(x)]^{q-1}
- \frac{q}{2} x^2 \xi^2 [f(x)]^{2(q-1)} +  o({{x^2\xi^2}
{[f(x)]^{2(q-1)}}}) \right)dx =$$
 \begin{equation}
 1 + i \xi \mu_{q} \nu_{q} - \frac{q}{2} \xi^2 \sigma^{2}_{2q-1} \nu_{2q-1}+
o(\xi^2 ), \, \xi \rightarrow 0.
\label{step_20}
  \end{equation}
In accordance with the condition of the theorem and the relation (2)
in Lemma \ref{qmeanproperties}, $\mu_{q}=0$ . Denote $Y_j =
{D_N(q)}^{-1}(X_j-\mu_{q}), j=1,2,...$. Then $Z_N = Y_1 +...+Y_N.$
Further, it is readily seen that, for a given random variable $X$
and real $a>0$, there holds $F_q [aX](\xi)=F_q[X](a^{2-q} \xi)$. It
follows from this relation that $F_{q}(Y_1)=F_{q}[f] \Bigl( \frac{\xi}{
\sqrt{N \nu_{2q-1}}\sigma_{2q-1}  } \Bigr)$ . Moreover, it follows from
the $q$-independence of $X_1,X_2,...$ and the associativity of the
$q$-product that
\begin{equation}
\label{step100} F_{q}[Z_N](\xi)= F_{q}[f]( \frac{\xi}{ \sqrt{N
\nu_{2q-1}}\sigma_{2q-1}  } ) {\otimes_q ... \otimes_q} F_{q}[f](
\frac{\xi}{ \sqrt{N \nu_{2q-1}}\sigma_{2q-1} } )
\,\,(N\,\mbox{factors}).
\end{equation}
Hence, making use of properties of the $q$-logarithm, from (\ref{step100})
we obtain
\[
\ln_q F_{q}[Z_N](\xi)= N \ln_q F_{q}[f]( \frac{\xi}{ \sqrt{N
\nu_{2q-1}}\sigma_{2q-1}  } ) = N \ln_q ( 1- \frac{q}{2}
\frac{\xi^2}{N} + o(\frac{\xi^2}{N})) =
\]

\begin{equation}
\label{step101} -\frac{q}{2} \xi^2 + o(1), \, N \rightarrow \infty
\,,
\end{equation}
locally uniformly by $\xi$.
\par
Consequently, locally uniformly by $\xi,$
\begin{equation}
\label{step_50}
\lim_{N \rightarrow \infty} F_{q}(Z_N) =
e_q^{-\frac{q}{2} \xi^2}.
\end{equation}
Thus, $Z_N$ is $q$-convergent to the random variable $Z$ whose
$q$-Fourier transform is $e_q^{-\frac{q}{2} \xi^2} \in
\mathcal{G}_q$.
\par
In accordance with Corollary \ref{centrallemmacor} for $q$ and some
$\beta$ there exists a density $G_{q_{-1}}(\beta; x),$ $q =
z(q_{-1}),$ such that $F_{q_{-1}}(G_{q_{-1}}(\beta; x))= e_q^{-(q/2)
\xi^2}.$ Let us now find $\beta$. It follows from Corollary
\ref{centrallemmacor} (see (\ref{gausstransform22})) that
$\beta_{\ast}(q_{-1})=q/2.$ Solving this equation with respect to
$\beta$ we obtain
\begin{equation}
\label{beta} \beta= \left( \frac{3-q_{-1}}{4q
C^{2(q_{-1}-1)}_{q_{-1}}} \right)^{\frac{1}{2-q_{-1}}},
\end{equation}
The explicit form of the corresponding
$q_{-1}$-Gaussian reads
as
\begin{equation}
\label{explicitGauss} G_{q_{-1}}(\beta; x)=     C^{-1}_{s}   \left(
\frac{\sqrt{3-s}}{2 C^{s-1}_{s}\sqrt{z(s)}} \right)^{\frac{1}{2-s}}
e_{s}^{ -  \left( \frac{3-s}{4z(s)C_{s}^{2(s-1)}}
\right)^{\frac{1}{2-s}}      x^2}, \, \, s=q_{-1}.
\end{equation}
\end{proof}

Analogously, the q-CLT can be proved for $q_k$-i.i.d. of the second and third
types. The formulations of the corresponding theorems are
given below. The reader can readily verify their validity through comparison with
the proof of Theorem 1.

\vspace{.3cm}

{\bf Theorem 3.} {\it Assume a sequence $q_k, k \in \mathcal{Z},$ is
given as (\ref{qn}) with $q_{k} \in[1,2).$ Let $X_1, ..., X_N,...$
be a sequence of $q_k$-independent (for a fixed $k$) of second type
and identically distributed random variables with a finite
$q_{k-1}$-mean $\mu_{q_{k-1}}$ and a finite second
$(2q_{k-1}-1)$-moment $\sigma_{2q_{k-1}-1}^2$ .

Then $Z_N = \frac{X_1 + ... + X_N -N \mu_{q_{k-1}}}{D_{N}(q_{k-1})}$
is $q_{k-1}$-convergent to a $q_{k-1}$-normal distribution as $N
\rightarrow \infty$.  The parameter $\beta_k$ of the corresponding
attractor $G_{q_{k-1}}(\beta_k; x)$ is
\begin{equation}
\label{betak2} \beta_k = \Bigl(\frac{3-q_{k-1}}{4 q_{k-1}
C_{q_{k-1}}^{2 q_{k-1} -2}}\Bigr)^{1 \over {2-q_{k-1}}}.
\end{equation}}

\vspace{.3cm}

{\bf Theorem 4.} {\it Assume a sequence $q_k, k \in \mathcal{Z},$ is
given as (\ref{qn}) with $q_{k} \in[1,2).$ Let $X_1, ..., X_N,...$
be a sequence of $q_k$-independent (for a fixed $k$) of third type
and identically distributed random variables with a finite
$q_{k}$-mean $\mu_{q_{k}}$ and a finite second $(2q_{k}-1)$-moment
$\sigma_{2q_{k}-1}^2$ .

Then $Z_N = \frac{X_1 + ... + X_N -N \mu_{q_{k}}}{D_{N}(q_{k})}$
is $q_{k-1}$-convergent to a $q_{k-1}$-normal distribution as $N
\rightarrow \infty$.  Moreover, the corresponding attractor
$G_{q_{k-1}}(\beta_k; x)$ in this case is the same as in the Theorem
1 with $\beta_k$ given in (\ref{betak}).}

\vspace{.3cm}

Obviously, $\frac{q+1}{3-q}=1$ if and only if $q=1.$ This fact yields the following corollary.
\begin{cor}
\label{thmcor} Let $X_1, ..., X_N,...$ be a given sequence of
$q_k$-independent (of any type) random variables satisfying the
corresponding conditions of Theorems 1-3.
Then the attractor of $Z_N$ is a $q_k$-normal distribution if and
only if $q_k=1,$ that is, in the classic case.
\end{cor}
\section{CONCLUSION}

In the present paper we studied $q$-generalizations of the classic
central limit theorem adapted to nonextensive statistical
mechanics, depending on three types of correlation. Interrelation
showing the dependence of a type of convergence to a $q$-Gaussian on
a type of correlation is summarized in Table 1.

\begin{table}[h]
{\small \centering \caption{Interrelation between the type of
correlation, conditions for $q$-mean and $q$-variance, type of
convergence, and parameter of the attractor }

\vspace{.5cm}

\begin{tabular}{|l|l|l|l|l|} \hline

{\bf Corr. type} & {\bf Conditions} & {\bf Convergence} & {\bf
Gaussian parameter } \\ \hline

       {1st type  } &
        {$\mu_{q_k}< \infty,$ \, $\sigma_{2q_k-1}^2 < \infty$} &
       { $q_{k}-conv.$ } &
       { $\beta_k = \Bigl(\frac{3-q_{k-1}}{4 q_{k}
C_{q_{k-1}}^{2 q_{k-1} -2}}\Bigr)^{1 \over {2-q_{k-1}}}$ } \\

\hline

         {2nd type  } &
{$\mu_{q_{k-1}}< \infty,$ } &
         { $q_{k-1}-conv.$ } &
         {$\beta_k = \Bigl(\frac{3-q_{k-1}}{4 q_{k-1}
C_{q_{k-1}}^{2 q_{k-1} -2}}\Bigr)^{1 \over {2-q_{k-1}}}$  } \\
{} & {$\sigma_{2q_{k-1}-1}^2 < \infty$} & {}  & {} \\

\hline

         {3rd type  } &
         {$\mu_{q_k}< \infty,$ \, $\sigma_{2q_k-1}^2 < \infty$} &
         { $q_{k-1}-conv.$ } &
         {$\beta_k = \Bigl(\frac{3-q_{k-1}}{4 q_{k}
C_{q_{k-1}}^{2 q_{k-1} -2}}\Bigr)^{1 \over {2-q_{k-1}}}$  } \\
\hline

\end{tabular}
\label{table} }
\end{table}

In all three cases the corresponding attractor is distributed
according to a $q_{k-1}$-normal distribution, where
$q_{k-1}=z^{-1}(q_k)$. We have noticed that $q_k \neq q_{k-1}$ if
$q_k \neq 1.$ In the classic case both $q_k=1$ and $z^{-1}(q_k)=1$,
so that the corresponding Gaussians do not differ. So, Corollary
\ref{thmcor} notes that such duality is a specific feature of the
nonextensive statistical theory, which comes from specific
correlations of $X_j$, which cause in turn nonextensivity of the phenomenon under
study.
\par
Now let us briefly discuss the scaling rate, important notion in
diffusion theory. We recall that the standard Gaussian evolved in
time can be calculated
$$G(t,x)=F^{-1}[e^{-t\xi^2}]=\frac{1}{2\sqrt{\pi
t}}e^{-\frac{x^2}{4t}}, \, t>0.$$ It follows immediately that the
mean squared displacement is related to time with the scaling rate $x^2 \sim t^\delta$ with
$\delta =1.$ Applying the same technique in the case of the $q$-theory,
in all three types of correlations, using (\ref{explicitGauss}),
Lemma \ref{qnrelation} and expression for $\beta_k$ (see fourth
column of Table 1), we obtain $\delta=1/(2-q_{k-1})=q_{k+1}.$ Hence,
three consequent members of $q_k=\frac{2q+k(1-q)}{2+k(1-q)}, \, q
\in [1, \, 2)$, namely the triplet $(q_{k-1},q_k,q_{k+1})$, play an
important role in the description of a nonextensive phenomenon. Namely, if
a correlation is given by $q_k$, then the corresponding attractor is
a $q_{k-1}$-Gaussian and, in turn, the scaling rate is equal to
$q_{k+1}.$
\par
Finally, we would like to address connections of the theorems that we have proved
with two known results. We have seen that the classical central
limit theorem may in principle be generalized in various manners,
each of them referring to correlations of specific kinds. In
\cite{MoyanoTsallisGellmann2006}, an example of correlated
model was discussed numerically. The correlations were introduced,
in a scale-invariant manner, through a $q$-product in the space of
the joint probabilities of $N$ binary variables, with $0 \le q \le
1 \,$\footnote{Even though we are analyzing here the case $q \geq 1$, it is useful to compare our model with the one in
\cite{MoyanoTsallisGellmann2006}.}. It was numerically shown that
the attractors are very close (although not exactly \cite{Hilhorst}) to (double-branched) $Q$-Gaussians, with
$Q=2-\frac{1}{q} \in (-\infty,1]$, and that the model is
superdiffusive \cite{MarshFuentesMoyanoTsallis} with $1 \le \delta
\le 2$). The relation $Q=2-\frac{1}{q}$ corresponds to the
particular case $k=-1$ of the present Theorem 1. It comes from Lemma
\ref{qnrelation}, with $q_{-2}=2-\frac{1}{q_0}=2-\frac{1}{q}$, which
holds when k=-1. It should be noted the following connection between
these two models, related to the behavior of the scaling rate
$\delta$.
In the model introduced in \cite{MoyanoTsallisGellmann2006},
superdiffusion occurs with $\delta$ monotonically decreasing from
$2$ to $1$ when $q$ increases from $0$ to $1$
\cite{MarshFuentesMoyanoTsallis}\,. In our $k=-1$ model we have
$\delta=q \in [1, 3)$, that is, monotonically increasing from $1$ to
$3$, when $q_{-1}$ increases from $1$ to $2.$ The scaling rates in
these two models behave as a sort of continuation of each other, through
$q=1,$ on the interval $q \in [0,3),$ and in both models we observe only
superdiffusion.
\par
Another example is suggested by the exact stable solutions of a
nonlinear Fokker-Planck equation in \cite{TsallisBukman}. The
correlations\footnote{The correlation defined in
\cite{TsallisBukman} is different from the $q$-independence of types 1-3
introduced in this paper.} are introduced through a $q=2-Q$ exponent
in the spatial member of the equation (the second derivative term).
The solutions are $Q$-Gaussians with $Q \in (-\infty,3)$, and
$\delta=2/(3-Q) \in [0,\infty]$, hence both superdiffusion and
subdiffusion can exist in addition to normal diffusion. This model
is particularly interesting because the scaling $\delta=2/(3-Q)$ was
conjectured in \cite{tsallisangra}, and it was verified in various
experimental and computational studies
\cite{arpita,daniels,rapisardaEPN,maza}.

In the particular case of Theorem 1, $k=1$, we have
$\delta=q_2=1/(2-q)$. This result coincides with that of the nonlinear
Fokker-Planck equation mentioned above. Indeed, in our theorem
($k=1$) we require the finitness of $(2q-1)$-variance. Denoting
$2q-1=Q$, we get $\delta=1/(2-q)=2/(3-Q).$ Notice, however, that
this example differs from the nonlinear Fokker-Planck above. Indeed,
although we do obtain, from the finiteness of the second moment, the
same expression for $\delta$, the attractor is not a $Q$-Gaussian,
but rather a $\frac{Q+1}{2}$ -Gaussian.
\par
Summarizing, the present Theorems 1-4 suggest a quite general and
rich structure at the basis of nonextensive statistical mechanics.
Moreover, they recover, as particular instances, central relations
emerging in the above two examples. The structure we have presently
shown might pave a deep understanding of the so-called $q$-triplet
$(q_{sen},q_{rel},q_{stat})$, where {\it sen}, {\it rel} and {\it
stat} respectively stand for {\it sensitivity to the initial
conditions}, {\it relaxation}, and {\it stationary state}
\cite{Tsallistriplet,BurlagaVinas,BurlagaVinas2} in nonextensive statistics. This
remains however as a challenge at the present stage. Another open question -- very relevant in what concerns physical applications -- refers to whether the $q$-independence addressed here reflects a sort of asymptotic scale-invariance as $N$ increases.

\subsection*{Acknowledgments}We acknowledge thoughtful remarks by K. Chow, M.G. Hahn, R. Hersh, J.A. Marsh, R.S. Mendes, L.G. Moyano, S.M.D.
Queiros and W. Thistleton.
One of us (CT) has benefited from lengthy conversations on the subject with M. Gell-Mann and D. Prato.
Financial support by the Fullbright Foundation, NIH grant P20 GMO67594, SI International and
AFRL (USA), and by CNPq and Faperj (Brazil) is acknowledged as well.


\begin{thebibliography}{9999}






\bibitem{Tsallis1988}C. Tsallis, {\it Possible generalization of Boltzmann-Gibbs statistics},  J. Stat. Phys. {\bf 52}, 479 (1988). See also E.M.F. Curado and C. Tsallis, {\it Generalized statistical mechanics: connection with thermodynamics}, J. Phys. A {\bf 24}, L69
(1991) [Corrigenda: {\bf 24}, 3187 (1991) and {\bf 25}, 1019 (1992)], and C. Tsallis, R.S. Mendes and A.R. Plastino, {\it The role of constraints within generalized nonextensive statistics}, Physica A {\bf 261}, 534 (1998).


\bibitem{Tsallis2005}C. Tsallis, {\it Nonextensive statistical mechanics, anomalous diffusion and central limit theorems}, Milan Journal of
Mathematics {\bf 73}, 145  (2005).



\bibitem{PratoTsallis1999}D. Prato and C. Tsallis, {\it Nonextensive
 foundation of Levy distributions}, Phys. Rev. E {\bf 60}, 2398 (1999),
and references therein.


\bibitem{MoyanoTsallisGellmann2006}L.G. Moyano, C. Tsallis and M.
Gell-Mann, {\it Numerical indications of a $q$-generalised central limit
theorem}, Europhys. Lett. {\bf 73}, 813 (2006).





\bibitem{variousCLT}G. Jona-Lasinio, {\it The renormalization group: A probabilistic view}, Nuovo Cimento B {\bf 26}, 99 (1975), and {\it Renormalization group and probability theory}, Phys. Rep. {\bf 352}, 439 (2001), and references therein; P.A. Mello and B. Shapiro, {\it Existence of a limiting distribution for disordered electronic conductors}, Phys. Rev. B {\bf 37}, 5860 (1988); P.A. Mello and S. Tomsovic, {\it
Scattering approach to quantum electronic transport}, Phys. Rev. B {\bf 46}, 15963 (1992); M. Bologna, C. Tsallis and P. Grigolini,{\it Anomalous diffusion associated with nonlinear fractional derivative Fokker-Planck-like equation: Exact time-dependent solutions}, Phys. Rev. E {\bf 62}, 2213 (2000); C. Tsallis, C. Anteneodo, L. Borland and R. Osorio, {\it Nonextensive statistical mechanics and economics}, Physica A {\bf 324}, 89 (2003); C. Tsallis, {\it What should a statistical mechanics satisfy to reflect nature?}, in {\it Anomalous Distributions, Nonlinear Dynamics and Nonextensivity}, eds. H.L. Swinney and C. Tsallis, Physica D {\bf 193}, 3 (2004); C. Anteneodo, {\it Non-extensive random walks}, Physica A {\bf 358}, 289 (2005); S. Umarov and R. Gorenflo, {\it On multi-dimensional symmetric random walk models approximating fractional diffusion processes},  Fractional Calculus and Applied Analysis {\bf 8}, 73-88 (2005);
S. Umarov and S. Steinberg, {\it Random walk models associated with distributed fractional order differential equations}, to appear in IMS Lecture Notes -
Monograph Series;
F. Baldovin and A. Stella, {\it Central limit theorem for anomalous scaling due to correlations}, Phys. Rev. E {\bf 75}, 020101 (2007);
C. Tsallis, {\it On the extensivity of the entropy $S_q$, the $q$-generalized central limit theorem and the $q$-triplet}, in Proc. {\it International Conference on Complexity and Nonextensivity: New Trends in Statistical Mechanics} (Yukawa Institute for Theoretical Physics, Kyoto, 14-18 March 2005), eds. S. Abe, M. Sakagami and N. Suzuki, Prog. Theor. Phys. Supplement {\bf 162}, 1 (2006); D. Sornette, {\it Critical Phenomena in Natural Sciences} (Springer, Berlin, 2001), page 36.





\bibitem{TsallisBukman}C. Tsallis and D.J. Bukman, {\it Anomalous diffusion in the presence of external forces: exact time-dependent solutions and their thermostatistical basis}, Phys. Rev. E {\bf 54}, R2197 (1996).





\bibitem{Yoshihara1992}K. Yoshihara, {\it Weakly dependent stochastic sequences and their applications}, V.1. {\it Summation theory for weakly dependent sequences} (Sanseido, Tokyo, 1992).





\bibitem{Peligrad1986}M. Peligrad, {\it Recent advances in the central theorem and its weak invariance principle for mixing sequences of random variables (a survey)},
in {\it Dependence in probability and statistics}, eds. E. Eberlein and M.S.Taqqu,  Progress in Probability and Statistics {\bf 11}, 193 (Birkh\"aser, Boston, 1986).





\bibitem{Rio2000}E. Rio, {\it Theorie asymptotique des processus aleatoires faiblement dependants}, Mathematiques et Applications {\bf 31} (Springer, Berlin, 2000).





\bibitem{Doukhan1994}P. Doukhan, {\it Mixing properties and examples}, Lecture Notes in Statistics {\bf 85} (1994).





\bibitem{DehlingDenkerPhilipp1986}H. Dehling, M. Denker and W. Philipp, {\it Central limit theorem for mixing sequences of random variables under minimal condition}, Annals of Probability {\bf 14} (4), 1359 (1986).





\bibitem{Bradley2003}R.C. Bradley, {\it Introduction to strong mixing conditions}, V I,II, Technical report, Department of Mathematics, Indiana University, Bloomington (Custom Publishing of IU, 2002-2003).





\bibitem{DehlingMikoschSorensen2002}H.G. Dehling, T. Mikosch and M. Sorensen, eds., {\it Empirical process techniques for dependent data} (Birkh\"aser, Boston-Basel-Berlin, 2002).





\bibitem{Euler} L. Euler, {\it Introductio in Analysin Infinitorum}, T. 1, Chapter XVI, p. 259, Lausanne, 1748.






\bibitem{Jackson} F. H. Jackson, {\it  On q-Functions and a Certain Difference Operator}, Trans. Roy Soc. Edin. 46
(1908), 253­281.






\bibitem{Ernst} T. Ernst, {\it A method for q-calculus}, Journal of Nonlinear Mathematical Physics Volume 10, Number 4 (2003), 487­525






\bibitem{TsallisGellmannSato}C. Tsallis, M. Gell-Mann and Y. Sato, {\it Asymptotically scale-invariant occupancy of phase space makes the entropy $S_q$ extensive}, Proc. Natl. Acad. Sc. USA {\bf 102}, 15377 (2005).





\bibitem{TsallisQuimicaNova}C. Tsallis, {\it What are the numbers that experiments provide ?}, Quimica Nova {\bf 17}, 468 (1994).





\bibitem{GellmannTsallis}M. Gell-Mann and C. Tsallis, eds., {\it Nonextensive Entropy - Interdisciplinary Applications} (Oxford University Press, New York, 2004).





\bibitem{qnivanen}L. Nivanen, A. Le Mehaute and Q.A. Wang, {\it Generalized algebra within a nonextensive statistics}, Rep. Math. Phys. {\bf 52}, 437 (2003).





\bibitem{qborges}E.P. Borges, {\it A possible deformed algebra and calculus inspired in nonextensive thermostatistics}, Physica A {\bf 340}, 95 (2004).





\bibitem{Billingsley} P. Billingsley. {\it Probability and Measure.} John Wiley and Sons, 1995.




\bibitem{Durrett} R. Durrett. {\it Probability: Theory and Examples.} Thomson, 2005.





\bibitem{MarshallEarl}J. Marsh and S. Earl, {\it New solutions to scale-invariant phase-space occupancy for the generalized entropy $S_q$}, Phys. Lett. A {\bf 349}, 146-152  (2005).




\bibitem{UmarovTsallis2007} S. Umarov and C. Tsallis, {\it Multivariate Generalizations of the $q$-Central Limit Theorem}, preprint (2007) [cond-mat/0703533].




\bibitem{tsallisEPN}C. Tsallis, M. Gell-Mann and Y. Sato, {\it Extensivity and entropy production},  Europhysics News {\bf 36}, 186 (2005).




\bibitem{Hilhorst}H.J. Hilhorst and G. Schehr, {\it A note on $q$-Gaussians and non-Gaussians in statistical mechanics},        J. Stat. Mech. P06003 (2007).




\bibitem{MarshFuentesMoyanoTsallis}J.A. Marsh, M.A. Fuentes, L.G. Moyano and C. Tsallis, {\it Influence of global correlations on central limit theorems and entropic extensivity}, Physica A {\bf 372}, 183-202 (2006).





\bibitem{CarusoTsallis}F. Caruso and C. Tsallis, {\it Extensive $q$-entropy in quantum magnetic systems}, preprint (2006) [cond-mat/0612032].





\bibitem{Beckbook}C. Beck and F. Schlogel, {\it Thermodynamics of Chaotic Systems: An Introduction} (Cambridge University Press, Cambridge, 1993).





\bibitem{tsallisangra}C. Tsallis, {\it Some thoughts on theoretical physics}, Physica A {\bf 344}, 718 (2004).





\bibitem{arpita}A. Upadhyaya, J.-P. Rieu, J.A. Glazier and Y. Sawada, {\it Anomalous diffusion and non-Gaussian velocity distribution of Hydra cells in cellular aggregates}, Physica A {\bf 293}, 549 (2001).





\bibitem{daniels}K.E. Daniels, C. Beck and E. Bodenschatz, {\it Defect turbulence and generalized statistical mechanics}, in {\it Anomalous Distributions, Nonlinear Dynamics and Nonextensivity}, eds. H.L. Swinney and C. Tsallis, Physica D {\bf 193}, 208 (2004).





\bibitem{rapisardaEPN}A. Rapisarda and A. Pluchino, {\it Nonextensive thermodynamics and glassy behavior}, Europhysics News {\bf 36}, 202 (2005).




\bibitem{maza}R. Arevalo, A. Garcimartin and D. Maza, {\it Anomalous diffusion in silo drainage}, Eur. Phys. J. E {\bf 23}, 191-198 (2007).




\bibitem{Tsallistriplet}C. Tsallis, {\it Dynamical scenario for nonextensive statistical mechanics}, in {\it News and Expectations in Thermostatistics}, eds. G. Kaniadakis and M. Lissia, Physica A {\bf 340}, 1 (2004).





\bibitem{BurlagaVinas}L.F. Burlaga and A.F. Vinas, {\it Triangle for the entropic index $q$ of non-extensive statistical mechanics observed by Voyager 1 in the distant heliosphere}, Physica A {\bf 356}, 375 (2005).




\bibitem{BurlagaVinas2}L.F. Burlaga, A.F. Vinas, N.F. Ness and M.H. Acuna, {\it Tsallis statistics of magnetic field in heliosheath}, Astrophys. J. Lett. {\bf 644}, L83-L86 (2006).



\end{thebibliography}
\end{document}